\documentclass[preprint,showpacs,preprintnumbers,amsmath,amssymb]{revtex4}

\usepackage{graphicx}
\usepackage{dcolumn}
\usepackage{bm}

\begin{document}


\title{A Supersymmetry Model of Leptons} 

\author{Chun Liu}  
\affiliation{Institute of Theoretical Physics, 
Chinese Academy of Sciences,\\
P. O. Box 2735, Beijing 100080, China}
 \email{liuc@itp.ac.cn}

\date{\today}

\begin{abstract}
  If supersymmetry (SUSY) is not for stabilizing the electroweak 
energy scale, what is it used for in particle physics?  We propose 
that it is for flavor problems.  A cyclic family symmetry is 
introduced.  Under the family symmetry, only the $\tau$-lepton is 
massive due to the vacuum expectation value (VEV) of the Higgs field.  
This symmetry is broken by a sneutrino VEV which results in the muon 
mass.  The comparatively large sneutrino VEV does not result in a 
large neutrino mass due to requiring heavy gauginos.  SUSY breaks 
at a high scale $\sim 10^{13}$ GeV.  The electroweak energy scale is 
unnaturally small.  No additional global symmetry, like the R-parity, 
is imposed.  Other aspects of the model are discussed.  
\end{abstract}



\maketitle

In elementary particle physics, SUSY \cite{susy} was proposed for 
stabilizing the electroweak (EW) scale \cite{susy1} which is otherwise 
unnaturally small compared to the grand unification scale \cite{gut}.  
The study of the cosmological constant \cite{cc}, however, suggests 
that unnaturalness of $10^{120}$ or $10^{55}$ fine tuning might be 
just so from the anthropic point of view.  It was argued that the 
string theory even supports the emergence of the anthropic landscape 
\cite{string}.  This led to a consideration of giving up naturalness 
of the EW scale \cite{split,split1}.  If SUSY is not for stabilizing 
the EW scale, what else job does it do in particle physics?  Refs. 
\cite{split,split1} maintained its roles in grand unification and the 
dark matter.  

In this paper, we advocate that SUSY is for flavor physics.  The 
flavor puzzle, namely the fermion masses, mixing and CP violation, in 
the Standard Model (SM) needs new physics to be understood.  The 
empirical fermion mass pattern is that the third generation is much 
heavier than the second generation which is also much heavier than the 
first.  This may imply a family symmetry 
\cite{family-symmetry,liu1,liu2}.  Let us consider the charged leptons.  
By assuming a $Z_3$ cyclic symmetry among the $SU(2)$ doublets 
$L_i$ ($i=1,2,3$) of the three generations \cite{liu1,liu2}, the 
Yukawa interactions result in a democratic mass matrix which is of 
rank $1$.  Therefore only the tau lepton gets mass, the muon and 
electron are still massless.  

The essential point is how the family symmetry breaks.  Naively the 
symmetry breaking can be achieved by introducing family-dependent 
Higgs fields.  We consider this problem within SUSY.  We observe that 
SUSY naturally provides such Higgs-like fields, which are the scalar 
neutrinos.  If the VEVs of the sneutrinos 
are non-vanishing, $v_i\neq 0$, the R-parity violating interactions 
$L_iL_jE_k^c$ \cite{r-parity}, with $E_k^c$ denoting the anti-particle 
superfields of the SU(2) singlet leptons, contribute to the fermion 
masses, in addition to the Yukawa interactions.  We think that this is 
the origin of family symmetry breaking.  

The above idea has been proposed for some time \cite{liu1,liu2}.  
Because SUSY was used to stabilize the EW scale, that idea suffers 
from severe constraints.  For example, the $\tau$-neutrino should be 
$10$ MeV heavy \cite{liu3}.  It is a liberation if SUSY has nothing to 
do with the EW scale.  While the $\tau$-lepton mass is from the Higgs 
VEV $\sim 100$ GeV, the $\mu$ mass is due to $v_i$, 
$m_\mu\sim\lambda v_i$ with $\lambda$ standing for the trilinear 
R-parity violation couplings.  It is natural $\lambda\sim 10^{-2}$ 
like the Yukawa couplings for the $\tau$ mass.  The muon mass tells us 
then $v_i\sim 10$ GeV.  $10$ GeV $v_i$'s could induce a large lepton 
number violating effect, namely a large neutrino Majorana mass if the 
neutralinos are not heavy, due to 
$\displaystyle m_\nu\simeq (g_2 v_i)^2/M_{\tilde{Z}}$, where $g_2$ is 
the $SU(2)_L$ gauge coupling constant, and $M_{\tilde{Z}}$ is the 
gaugino mass.  When we get the freedom to take $M_{\tilde{Z}}$ 
arbitrarily high, the above formula can produce a neutrino mass in the 
safe range.  

In this model the $Z_{3L}$ family symmetry mentioned above is assumed, 
which however is softly broken.  The gauge symmetries and the matter 
contents in the full theory are the same as those in the SUSY SM.  
Under the family symmetry, the relevant kinetic terms generally 
include 
\begin{equation}
\label{1}
\begin{array}{lll}
{\mathcal  L} &\supset& 
\left( H_1^{\dag}H_1+H_2^{\dag}H_2+\alpha L_i^{\dag}L_i
+\beta (L_1^{\dag}L_2+L_2^{\dag}L_3+L_3^{\dag}L_1+h.c.)\right.\\[3mm]
 & &\displaystyle 
+\left.\left. \frac{\gamma}{\sqrt{3}}(H_2^{\dag}\sum_i L_i+h.c.)\right)
\right|_{\theta\theta\bar{\theta}\bar{\theta}} \,,
\end{array}
\end{equation}
where $H_1$ and $H_2$ are the two Higgs doublets, $\alpha$, $\beta$, 
$\gamma$ are $O(1)$ coefficients.  The case of that $\alpha=1$ and 
$\beta=\gamma=0$ is a special one of above expression.  Note that the 
gauge field $e^V$ is not explicitly written, which does not affect our 
discussion on flavor physics.  The superpotential is 
\begin{equation}
\label{2}
{\mathcal W} = \frac{\tilde{y}_j}{\sqrt{3}}(\sum_i L_i)H_2E^c_j 
+\tilde{\lambda}_j(L_1L_2+L_2L_3+L_3L_1)E^c_j
+\tilde{\mu}H_1H_2+\tilde{\mu}'H_1\sum_i L_i\,,
\end{equation}
where $\tilde{y}_j$'s and $\tilde{\lambda}_j$'s are the coupling 
constants.  $\tilde{\mu}$ and $\tilde{\mu}'$ are mass terms.  It is 
natural that they are about the scale of soft SUSY breaking masses.  
The Lagrangian of soft SUSY breaking masses is 
\begin{equation}
\label{3}
\begin{array}{lll}
{\mathcal L}_{soft1}&=& M_{\tilde{W}}\tilde{W}\tilde{W}
             +M_{\tilde{Z}}\tilde{Z}\tilde{Z} \\[3mm] 
&&+m_h^2h_1^{\dag}h_1+m_h^2h_2^{\dag}h_2
+m_{lL_{ij}}^2\tilde{l}_i^{\dag}\tilde{l}_j
+m_{lR_{ij}}^2\tilde{e}_i^*\tilde{e}_j \\[3mm] 
&&+(B_{\tilde{\mu}} h_1h_2+B_{\tilde{\mu}_i} h_1\tilde{l_i}
+m_i^{\prime 2} h_2^{\dag}\tilde{l}_i+h.c.)\,,
\end{array}
\end{equation}
where $\tilde{W}$ and $\tilde{Z}$ stand for the charged and neutral 
gauginos, respectively,  $h_1$, $h_2$, $\tilde{l}_i$ and $\tilde{e}_i$ 
are the scalar components of $H_1$, $H_2$, $L_i$ and $E^c_i$ 
respectively.  Note that explicitly breaking of $Z_{3L}$ is introduced 
in the soft mass terms.   The soft masses are assumed to be very large 
around a typical mass $m_S$.  The trilinear soft terms should be also 
included, 
\begin{equation}
\label{4}
{\mathcal L}_{soft2}=\tilde{m}_{ij}\tilde{l}_ih_2\tilde{e}_j
+\tilde{m}_{ijk}\tilde{l}_i\tilde{l}_j\tilde{e}_k+h.c.\,.  
\end{equation}
The mass coefficients which we denote generally as $\tilde{m}_S$ can 
be close to $m_S$.  

The expression of the kinetic terms is not yet in the normalized 
standard form.  The standard form 
\begin{equation}
\label{5}
{\mathcal L} \supset H_u^{\dag} H_u+H_d^{\prime\dag}H_d'+L_e^{\dag}L_e
+L_{\mu}^{\dag}L_{\mu}+L_{\tau}^{\prime\dag}L_{\tau}'   
\end{equation}
is achieved by the field re-definition: 
\begin{equation}
\label{6}
\begin{array}{lll}
H_u    &=& H_1 \\
H_d'   &=& \displaystyle 
c_1 \left(H_2+\frac{c_2}{\sqrt{3}}\sum_i L_i\right) \\[3mm]
L_{\tau}'&=& \displaystyle 
c_1'\left(H_2-\frac{c_2}{\sqrt{3}}\sum_i L_i\right) \\[3mm]
L_{\mu}  &=& \displaystyle \frac{c_3}{\sqrt{2}}(L_1-L_2)\cos\theta
          +\frac{c_3}{\sqrt{6}}(L_1+L_2-2L_3)\sin\theta        \\[3mm]
L_e    &=&-\displaystyle \frac{c_3}{\sqrt{2}}(L_1-L_2)\sin\theta
          +\frac{c_3}{\sqrt{6}}(L_1+L_2-2L_3)\cos\theta \,,
\end{array}
\end{equation}
where 
\begin{equation}
\label{7}
\begin{array}{ll}
c_1 &=\displaystyle\frac{1}{\sqrt{2}}\sqrt{1+\frac{\gamma}{c_2}}\, ~~
     c_2=\sqrt{\alpha+2\beta}\, ~~ c_3=\sqrt{\alpha-\beta}\, \\[3mm]
c_1'&=\displaystyle \frac{1}{\sqrt{2}}\sqrt{1-\frac{\gamma}{c_2}}\,
\end{array}
\end{equation}
and $\theta$ can not be determined until muon mass basis is fixed.  

The superpotential is then 
\begin{equation}
\label{8}
\begin{array}{lll}
{\mathcal W} &=& \sqrt{\sum_j |y_j|^2} H_d'L_{\tau}' E^c_{\tau}  
+L_eL_{\mu}(\lambda_{\tau} E^c_{\tau}+\lambda_{\mu} E^c_{\mu}) \\[3mm] 
   & & +\mu H_uH_d'+ \mu' H_u L_{\tau}'\,,
\end{array}
\end{equation}
where 
\begin{equation}
\label{9}
\begin{array}{llll}
y_j & = & \displaystyle\frac{2}{\sqrt{\alpha+2\beta-\gamma^2}}
          \tilde{y_j}\,, ~~& 
\lambda_j = \displaystyle -\frac{\sqrt{3}}{\alpha+\beta}
\tilde{\lambda_j}\,,\\[3mm]
\mu & = & \displaystyle \frac{1}{2c_1}\left(\tilde{\mu}
+\frac{\tilde{\mu}'}{c_2}\right)\,, ~~ &
\mu' = \displaystyle \frac{1}{2c_1'}\left(\tilde{\mu}
-\frac{\tilde{\mu}'}{c_2}\right)\,,
\end{array}
\end{equation}
$E^c_{\tau}$ is defined as 
\begin{equation}
\label{10}
E^c_{\tau} = \frac{1}{\sqrt{\sum_j |y_j|^2}} y_jE^c_j\,, 
\end{equation} 
$E^c_{\mu}$ is orthogonal to $E^c_{\tau}$, $\lambda_{\tau}$ and 
$\lambda_{\mu}$ are combinations of $y_j$'s and $\lambda_j$'s.  Because 
of the $Z_{3L}$ symmetry, the superpotential is without the field 
$E^c_e$ which is orthogonal to both $E^c_{\tau}$ and $E^c_{\mu}$.  

To look at the fermion masses, we simply rotate the bilinear R-parity 
violating term away via the field re-definition, 
\begin{equation}
\label{11}
\begin{array}{lll}
H_d &=& \displaystyle\frac{1}{\sqrt{\mu^2+\mu'^2}}
        (\mu H_d'+\mu'L_{\tau}')\,, \\[3mm] 
L_{\tau} &=& \displaystyle 
           \frac{1}{\sqrt{\mu^2+\mu'^2}}(\mu' H_d'-\mu L_{\tau'})\,.  
\end{array}
\end{equation}
It is trivial to see that the kinetic terms are diagonal in terms of 
$H_d$ and $L_{\tau}$.  The superpotential is 
\begin{equation}
\label{12}
\begin{array}{lll}
{\mathcal W} &=& -\sqrt{\sum_j |y_j|^2} H_d L_{\tau} E^c_{\tau}  
+L_eL_{\mu}(\lambda_{\tau} E^c_{\tau}+\lambda_{\mu} E^c_{\mu}) \\[3mm] 
   & & +\sqrt{\mu^2+\mu'^2} H_uH_d \,.  
\end{array}
\end{equation}
The $Z_{3L}$ family symmetry keeps the trilinear R-parity violating 
terms invariant.  As we have expected Higgs field $H_d$ contributes to 
the tauon mass only and the sneutrinos in $L_e$ and $L_{\mu}$ contribute 
to the muon mass, after they get VEVs.  The VEVs of $L_e$ and $L_{\mu}$ 
imply the breaking of the $Z_{3L}$ symmetry as can be seen explicitly 
from Eq. (\ref{6}).  The electron remains massless because of absence of 
the $E^c_e$ field in ${\mathcal  W}$.  A hierarchy among charged leptons 
is obtained.  Without losing our essential points, we could take 
$\lambda_{\tau}=0$.  In that case, $L_{\tau}$ is in the mass eigenstate.  
And the tau number is conserved.  The tauon number conservation 
justifies the field rotation Eq. (\ref{11}).

The breaking of the family symmetry originates from the soft SUSY 
masses.  For simplicity and without losing generality, we assume 
that the soft terms in Eqs. (\ref{3}) and (\ref{4}) are rewritten as 
\begin{equation}
\label{13}
\begin{array}{lll}
{\mathcal L}_{soft}&=& M_{\tilde{W}}\tilde{W}\tilde{W}
             +M_{\tilde{Z}}\tilde{Z}\tilde{Z} \\[3mm] 
&&+m_{h_u}^2h_u^{\dag}h_u+m_{h_d}^2h_d^{\dag}h_d
+m_{h_d}^2\tilde{l}_{\alpha}^{\dag}\tilde{l}_{\alpha}
+m_{lR_{\alpha\beta}}^2\tilde{e}^*_{\alpha} \tilde{e}_{\beta} \\[3mm] 
&&+(B_{\mu} h_uh_d+B_{\mu_e} h_u\tilde{l}_e \\[3mm]
&&+\tilde{m}_{\alpha\beta}\tilde{l}_{\alpha} h_d\tilde{e}_{\beta}
+\tilde{m}_{\alpha\beta\gamma}
\tilde{l}_{\alpha}\tilde{l}_{\beta}\tilde{e}_{\gamma}+h.c.)\,, 
\end{array}
\end{equation}
where $\alpha=e, \mu, \tau$.  Most of the squared masses are expected 
to be positive, except $m_{h_u}^2$.

The key point of the form of the soft masses lies in the 
$(h_u~~h_d^{\dag}~~\tilde{l}_e^{\dag})$ mass-squared matrix, 
\begin{equation}
\label{14}
{\mathcal M}^{(h_u,h_d^{\dag},\tilde{l}_e^{\dag})} = 
\left(
\begin{array}{ccc}
m_{h_u}^2     & B_{\mu}   & B_{\mu_e} \\
B_{\mu}   & m_{h_d}^2 & 0         \\
B_{\mu_e} & 0         & m_{h_d}^2 \\
\end{array}
\right) 
\end{equation}
of which the eigenvalues are 
\begin{equation}
\label{15}
\begin{array}{lll}
M_1^2 & = & \bar{m}-\sqrt{\Delta^2+(B_{\mu})^2+(B_{\mu_e})^2} \\
M_2^2 & = & \bar{m}+\sqrt{\Delta^2+(B_{\mu})^2+(B_{\mu_e})^2} \\
M_3^2 & = & m_{h_d}^2\,, 
\end{array}
\end{equation}
where $\bar{m}=\displaystyle\frac{m_{h_u}^2+m_{h_d}^2}{2}$, 
$\Delta=\displaystyle\frac{m_{h_u}^2-m_{h_d}^2}{2}$.  The analysis 
goes in the similar way as in Ref. \cite{split}.  By fine-tuning, 
$M_1^2\sim -m_{EW}^2$, namely the EW symmetry breaking is achieved.  
The tuning is at the order of $m_S^2/m_{EW}^2$.  

In our case, in addition to the Higgs doublets, $\tilde{l}_e$ field 
also gets a VEV, 
\begin{equation}
\label{16}
v_u \neq 0\,, ~~v_d\neq 0\,, ~~ v_{l_e} \neq 0\,.  
\end{equation}
The relative size of these values are determined by the soft mass 
parameters.  It is natural to expect the $Z_{3L}$ symmetry breaking is 
not large, a hierarchy between $v_{u,d}$ and $v_{l_e}$ is possible.  
In the extreme case of that $B_{\mu_e}\ll B_{\mu}$, $v_{l_e}$ vanishes.  
As an illustration, a prefered VEV pattern $v_u > v_d > v_{l_e}$ is 
expected if $|m_{h_u}^2| < |m_{h_d}^2|$ and $B_{\mu_e} < B_{\mu}$ are 
assumed.  Note that the $L_e$ number breaks explicitly in the soft mass terms, 
$v_{l_e}$ does not result in any massless scalar.  Because there is 
only one light Higgs doublet, the tree-level flavor changing neutral 
current (FCNC) does not appear.  Therefore, a vanishing $\lambda_{\tau}$ 
keeps generality of the model.  The fact  $v_{l_e}\neq 0$ results in 
$\cos\theta=1$, and it is at this stage that $E^c_{\mu}$ just 
corresponds to the mass eigenstate of the muon.  The hierarchical charged 
lepton mass pattern is obtained from Eq. (\ref{12}) explicitly, 
\begin{equation}
\label{17}
\begin{array}{lll}
m_{\tau} & = & \sqrt{\sum_j|y_j|^2} v_d \,, \\
m_{\mu}  & = & \lambda_{\mu} v_{l_e}      \,, \\ 
m_e      & = & 0                        \,. 
\end{array}
\end{equation}
Numerically it is required that $v_d\sim 100$ GeV and $v_{l_e}\sim 10$ 
GeV.  

Whethera a large $v_{l_e}$ is safe or not should be studied.  In addition, 
it should be also considered that a huge $B_{\mu_e}$ induces a large 
lepton-Higgsino mixing.  The inducement happens at the loop-level 
through the gaugino exchange, as shown in Ref. \cite{liu3}, 
$m_{eh}=\displaystyle\frac{g_2^2B_{\mu_e}}{16\pi^2M_{\tilde{Z}}}$ 
which is about $10^{-3}m_S$.  By denoting $\tilde{h}$ as Higgsinos, 
the mass matrix of $\nu_e$ and the other neutralinos is given as 
\begin{equation}
\label{18}
-i\left(\nu_e~\tilde{h}_d^0~\tilde{h}_u^0~\tilde{Z}\right)
\left(
\begin{array}{cccc}
0        & 0                    & m_{eh}               & av_{l_e} \\
0        & 0                    & -\sqrt{\mu^2+\mu'^2} & av_d     \\
m_{eh}   & -\sqrt{\mu^2+\mu'^2} & 0                    & -av_u    \\
av_{l_e} & av_d                 & -av_u                & M_{\tilde{Z}}
\end{array}
\right)
\left(
\begin{array}{c}
\nu_e \\ \tilde{h}_d^0 \\ \tilde{h}_u^0 \\ \tilde{Z} 
\end{array}
\right)\,,
\end{equation}
where $a=\displaystyle(\frac{g_2^2+g_1^2}{2})^{1/2}$ with $g_1$ being 
the SM $U(1)_Y$ coupling constant.  We simply obtain the mass 
eigenvalues (denoted as $\Lambda_1, \Lambda_2, \Lambda_3, \Lambda_4$) of the 
above mass matrix by reasonably taking 
$v_{l_e}\ll v_d < v_u\ll \sqrt{\mu^2+\mu'^2}\sim M_{\tilde{Z}}$, 
\begin{equation}
\label{19}
\begin{array}{lll}
\Lambda_1 & \simeq & M_{\tilde{Z}}                 \,,         \\
\Lambda_2 & \simeq & \sqrt{\mu^2+\mu'^2+m_{eh}^2}  \,,         \\
\Lambda_3 & \simeq & -\sqrt{\mu^2+\mu'^2+m_{eh}^2} \,,         \\[3mm] 
\Lambda_4 & \simeq & \displaystyle -\frac{(av_{l_e})^2}{M_{\tilde{Z}}} \,.
\end{array}
\end{equation} 
Therefore the $\nu_e$ mass 
$m_{\nu_e}\simeq\displaystyle\frac{(av_{l_e})^2}{M_{\tilde{Z}}}$.  It 
is very small $\sim 10^{-3}$ eV when $M_{\tilde{Z}}\sim 10^{13}$ GeV.  

To accommodate the neutrino oscillation data, the neutrino sector 
should be extended.  Three right-handed neutrinos $N_i~~(i=1,2,3)$ 
which are singlet under the SM gauge groups, are introduced.  The 
following terms should be included in the $Z_{3L}$ symmetric 
superpotential Eq. (\ref{2}), 
\begin{equation}
\label{20}
{\mathcal W} \supset \frac{y'_j}{\sqrt{3}}\sum_i L_i H_1N_j 
+M_{ij}N_iN_j+\tilde{c}_jH_1H_2N_j\,,
\end{equation}
with $y'_j$'s and $\tilde{c}_j$'s being the coupling constants 
of ${\mathcal O} (10^{-2})$, and $M_{ij}$ the Majorana masses.  
${\mathcal W}$ does not include purely linear terms of $N_i$'s with large 
mass-squared coefficients, because $N_i$'s are supposed to be charged 
under a larger gauge group beyond the SM.  The soft masses of $N_i$'s 
are simply assumed to be large enough that $N_i$'s do not develop 
non-vanishing VEVs.  The trilinear soft terms associated with $N_i$'s 
can be written explicitly, which however, play little roles in the 
analysis.  Through the previous field redefinition, 
Eq. (\ref{12}) then includes 
\begin{equation}
\label{21}
{\mathcal W} \supset y'_\tau H_u L_\tau N_\tau 
+M_{\alpha\beta}N_\alpha N_\beta 
+H_uH_d
(\tilde{c}_\tau N_\tau+\tilde{c}_\mu N_\mu+\tilde{c}_e N_e) \,,
\end{equation}
where $N_\alpha$'s are combinations of $N_i$'s with $N_\tau$ being 
that which couples to $H_u L_\tau$.  $y'_\tau$ and $\tilde{c}_\alpha$ 
are combinations of $y_i'$'s, $\tilde{c}_i$, $c_1^{(\prime)}$, $c_2$ 
and $\mu^{\prime}/\mu$.  The $\nu_\tau$ mass is determined by the see-saw 
mechanism from Eq. (\ref{21}), 
\begin{equation}
\label{22}
m_{\nu_{\tau}} \simeq \frac{(y'_{\tau} v_d)^2}{M_{\alpha\beta}} 
            \simeq 3\times 10^{-2} {\rm eV} 
\end{equation}
by taking $M_{\alpha\beta}\sim (10^{10}-10^{11})$ GeV.  The Dirac 
neutrino mass matrix is diagonal in the $e$-$\mu$-$\tau$ basis.  A   
bi-large neutrino mixing originates from the mass matrix 
$M_{\alpha\beta}$.  

The electron mass comes from the soft trilinear R-parity violating 
terms in Eq. (\ref{13}) \cite{liu1}.  Their soft breaking of $Z_{3L}$ 
generates non-vanishing masses for the charged leptons through the one 
loop diagram with a gaugino exchange.  The mixing of the scalar 
leptons associated with different chiralities is due to the soft 
trilinear terms, which is then about 
$\sqrt{\sum_j|y_j|^2}\tilde{m_S} v_d$.  The one loop contribution to the 
charged lepton masses is about 
\begin{equation}
\label{23}
\delta M^l_{\alpha\beta}\simeq \frac{\alpha}{\pi}
\frac{\sqrt{\sum_j|y_j|^2}\tilde{m_S} v_d}{m_S} \,.  
\end{equation}
Taking $\tilde{m_S}/m_S\simeq 0.1$, 
$\delta M^l_{\alpha\beta} \sim {\cal O}$(MeV) which determines the 
electron mass.  

The lepton mixing mainly depend on the neutrino mass matrix.  In the charged 
lepton mass matrix, $m_{\mu}$ and $m_{\tau}$ are at the diagonal positions, 
the non-diagonal elements are $\delta M^l_{\alpha\beta}$.  The mixing from 
the charged leptons are then basically small, 
$U_{\mu\tau}\simeq \displaystyle \sqrt{\frac{m_e}{m_{\tau}}}$, 
$U_{e\mu}\simeq \displaystyle \sqrt{\frac{m_e}{m_{\mu}}}$.  If the 
mixing due to the neutrino mass matrix is bi-large, the lepton mixing 
required by the neutrino oscillation data can be obtained.  

Let us briefly comment on the quark masses.  Like that of the charged 
leptons, the quark masses also have three origins: the Higgs VEVs, the 
sneutrino VEV and soft trilinear R-parity violating terms.  However, 
the roles of the sneutrino VEV and the soft trilinear terms are 
switched \cite{liu2}.  The sneutrino VEV contributes to the first 
generation quark masses, and the soft trilinear R-parity violating 
terms to the charm and strange quark masses.  More details will be in 
a separate work \cite{liu4}.  One important merit of this framework 
is that we do not need to introduce baryon number conservation.  Because 
the sparticles are very heavy, they suppress baryon number violating 
processes to be unobservable \cite{liu4}.  An essentially same 
observation was pointed out in split SUSY \cite{r-parity2}.  

In summary, we have proposed that SUSY is for flavor problems in 
particle physics.  A family symmetry $Z_{3L}$, which is the cyclic 
symmetry among the three generation $SU(2)_L$ doublets, is introduced.  
No R-parity is imposed.  SUSY breaks at a high scale 
$\sim 10^{13}$ GeV.  The electroweak energy scale is unnaturally 
small.  Under the family symmetry, only the $\tau$-lepton gets its 
mass.  This symmetry is broken by a sneutrino VEV which results in the 
muon mass.  A hierarchical pattern of the charged lepton masses are 
obtained.  The comparatively large sneutrino VEV does not result in a 
large neutrino mass because of the gaugino masses are very heavy.  
The quark masses and other aspects of the model have been also 
discussed.  

At low energies, the model is basically the same as the SM.   
One essential feature of this model is that the unnaturally light 
Higgs has a component of a slepton.  Related to this point, the model 
allows for relatively long-lived Higgsinos.  We may consider a case 
where their masses are lower than $m_S$.  If they are loop induced, 
the Higgsino masses are thousand times smaller than $m_S$.  A Higgsino 
decays to a Higgs and a virtual gaugino which further goes into a 
lepton and a virtual slepton, the slepton decays to a lepton pair via 
R-parity violating interaction.  Because this four body decay is 
suppressed by the R-parity violating coupling and double suppressed by 
$m_S$, a $10^{10}$ GeV heavy Higgsino has a lifetime of $10^{-12}$ 
sec.  The cosmological and astrophysical implications should be 
studied in future works.  

\begin{acknowledgments}
The author acknowledges support from the National Natural Science
Foundation of China.
\end{acknowledgments}


\newpage 

\begin{thebibliography}{99}

\bibitem{susy}
J. Wess and B. Zumino, Nucl. Phys. B 70 (1974) 39;\\
Y. Gol'fand and E. Likhtman, JETP Lett. 13 (1971) 323;\\
D.V. Volkov and V. Akulov, Phys. Lett. B 46 (1973) 109.

\bibitem{susy1}
E. Witten, Nucl. Phys. B 188 (1981) 513;\\
S. Dimopoulos and H. Georgi, Nucl. Phys. B 193, 150 (1981).  

\bibitem{gut}
J.C. Pati and A. Salam, Phys. Rev. D 10 (1974)275;\\
H. Georgi and S.L. Glashow, Phys. Rev. Lett. 32 (1974) 438.

\bibitem{cc}
S. Weinberg, Phys. Rev. Lett. 59, 2607 (1987).  

\bibitem{string}
R. Bousso and J. Polchinski, JHEP 0006, 006 (2000);\\
J. L. Feng, J. March-Russell, S. Sethi, F. Wilczek, 
Nucl. Phys. B 602, 307 (2001).

\bibitem{split}
N. Arkani-Hamed and S. Dimopoulos, hep-th/0405159.

\bibitem{split1}
G. F. Giudice and A. Romanino, Nucl. Phys. B 699, 65 (2004).  

\bibitem{family-symmetry}
H. Fritzsch, Phys. Lett. B 70, 436 (1977); 
S. Adler, Phys. Rev. D 59 (1999) 015012, Erratum D 59 (1999) 099902.  
For recent studies, see e.g. 
T. Kitabayashi and M. Yasu\'e, Phys. Rev. D 67 (2003) 015006; 
P. F. Harrison, W. G. Scott, Phys. Lett. B 557 (2003) 76.

\bibitem{liu1}
D. Du and C. Liu, Mod. Phys. Lett. A 8, 2271 (1993); 
A 10, 1837 (1995);\\ 
For a review, see C. Liu, in Beijing 1999, 
Frontier of Theoretical Physics, p. 131 [hep-ph/0005061].

\bibitem{liu2}
C. Liu, Int. J. Mod. Phys. A 11, 4307 (1996).  

\bibitem{r-parity}
C. Aulakh and R. Mohapatra, Phys. Lett. B 119, 136 (1982); \\
S. Weinberg, Phys. Rev. D 26, 287 (1982); \\
For reviews, see G. Bhattacharyya, hep-ph/9709395, 
O.C.W. Kong, Int. J. Mod. Phys. A 19, 1863 (2004). 

\bibitem{liu3}
C. Liu and H. S. Song, Nucl. Phys. B 545, 183 (1999).  

\bibitem{liu4}
C. Liu, in preparation.  

\bibitem{r-parity2}
S.K. Gupta, P. Konar, B. Mukhopadhyaya, hep-ph/0408296.  

\end{thebibliography}
\end{document}